\documentstyle[preprint,tighten,eqsecnum,aps,epsfig]{revtex}
\begin{document}
\preprint{TRI-PP-95-59}
\draft
\title{$\eta \to \pi^0 \gamma \gamma$ to ${\cal{O}}(p^6)$
in Chiral Perturbation Theory }
\author{M.\ Jetter\footnote{Address after August 31, 1995: M. Jetter, c./o.
Werner Jetter, Hertenwinkelstr.\ 15, \\ 72336 Balingen/Germany.
Fax: 07433-23173.} }
\address{TRIUMF, 4004 Wesbrook Mall, Vancouver, B.\ C., \\
 Canada  V6T 2A3}
\date{\today}
\maketitle
\begin{abstract}
The decay $\eta \to \pi^0 \gamma \gamma$ is discussed in the framework
of SU(3) chiral perturbation theory. The process is dominated by the
${\cal{O}}(p^6)$ in the momentum expansion where tree-level amplitudes from
the effective Lagrangian ${\cal{L}}^6$ enter together with one--loop
contributions from ${ \cal{L}}^4$ and two--loop contributions from
${\cal{L}}^2$.
We estimate the 6 independent ${\cal{L}}^6$ coupling constants by
resonance saturation consistent with the pion production process
$\gamma \gamma \to \pi^0 \pi^0$ and calculate the pion--loop part
of the one- and two--loop amplitude. Predictions for the total rate and
spectrum of $\eta \to \pi^0 \gamma \gamma$
are given together with a discussion of the uncertainties involved.
\end{abstract}
\narrowtext
\section{Introduction}

The reactions involving two neutral pseudoscalar mesons and two photons have
received much attention during the past few years. In the framework of
chiral perturbation theory, they are unique in that the tree--level
amplitudes from the Lagrangians ${\cal{L}}^2$ and ${\cal{L}}^4$ both vanish.
As a consequence, processes such as
$\gamma \gamma \to \pi^0 \pi^0$, $\eta \to \pi^0 \gamma \gamma$ (or the
reactions related to them by time reversal) test higher order and
loop contributions in the chiral perturbation series.
In fact it turns out that the ${\cal{O}}(p^4)$ one-loop amplitude for
$\gamma \gamma \to \pi^0 \pi^0$ \cite{Hol88,Bij88} underpredicts
the Crystal Ball data \cite{Data94} at low invariant mass of the two photons
(where chiral perturbation theory is supposed to work best). The
${\cal{O}}(p^4)$ result for the $\eta \to \pi^0 \gamma \gamma$ decay--width
is a tiny $3.89 \cdot 10^{-3}$ eV \cite{Bij92}, to be compared with an
experimental value of $0.84 \pm 0.2$ eV \cite{Data90}.
The $\eta$--decay amplitude is thus dominated by higher order contributions.

According to Weinberg's power--counting formula \cite{Wei79},
the ${\cal{O}}(p^6)$--amplitude
will contain tree-level contributions from the effective Lagrangian
${\cal{L}}^6$ together with one--loop
contributions from ${ \cal{L}}^4$ and two--loop contributions
from ${\cal{L}}^2$. The form of the complete ${\cal{L}}^6$--Lagrangian
has been given recently \cite{Sche94}; we will see below that
a total of 6 linearly independent structures contribute to the
processes considered here. In extending the chiral Lagrangian approach
to ${\cal{O}}(p^6)$, we therefore have to deal with two difficulties:
A realistic estimate of the 6 low--energy constants involved and the
solution of the 2--loop Feynman integrals, in particular for the case where
the two loop momenta overlap.

There have been various attempts to determine the ${\cal{L}}^6$
$\eta$--decay amplitude in the past either phenomenologically or in
a bosonized Nambu-Jona-Lasinio model. In \cite{Bij92}, only the leading
terms in a $1/N_c$--expansion are considered, i.\ e. the Lagrangian
is restricted to the single flavour trace terms
\begin{equation} d_1 F_{\mu \nu} F^{\mu \nu} Tr(Q^2 \delta^{\lambda} U
\delta_{\lambda}
U^{\dagger}) + d_2 F_{\mu \alpha} F^{\mu \beta} Tr(Q^2 \delta^{\alpha}
 U \delta_{\beta} U^{\dagger}). \label{Bijlag} \end{equation}
(For the notation, see Section \ref{Notation}).
If the two coupling constants $d_1$ and $d_2$ are determined by comparison
to a vector dominance (VMD) amplitude
$$ -d_1={1 \over 2} d_2 \sim 3.6 \cdot 10^{-3} {\rm GeV}^{-2},
\label{Bijconst}$$
the tree level amplitude from the ${\cal{O}}(p^6)$ Lagrangian (\ref{Bijlag})
yields a total decay width of
$$ \Gamma^6 (\eta \to \pi^0 \gamma \gamma) = 0.18 \, {\rm eV}.$$
This is to be compared to the full VMD amplitude of \cite{Bij92,Ng92,Pic92}
$$ \Gamma^{VMD} (\eta \to \pi^0 \gamma \gamma) \sim 0.31 \, {\rm eV}.$$

In the Nambu--Jona--Lasinio model, a further single trace term contributes
to ${\cal{L}}^6$, namely \cite{Belu95,Belk95}
\begin{equation}  d_3 F_{\mu \nu} F^{\mu \nu}
Tr(Q^2 (\chi^{\dagger} U + U^{\dagger} \chi)).
\label{njllag} \end{equation}
The determination of the constant $d_3$ is ambiguous. According to Ref.
\cite{Belu95}, the term (\ref{njllag}) increases the
$\eta$--decay width to
$$ \Gamma^{ENJL} (\eta \to \pi^0 \gamma \gamma) = 0.58 \pm 0.3 \, {\rm eV}$$
whereas the authors of \cite{Belk95} advocate values of
$$ \Gamma^{NJL} (\eta \to \pi^0 \gamma \gamma) = 0.11 \ldots 0.45
 \, {\rm eV}. $$
If $d_3$ is calculated via resonance saturation in
$\gamma \gamma \to \pi^0 \pi^0$
\cite{Gas94}, the term (\ref{njllag}) doesn't affect the $\eta$--decay width
very much. Conversely, an attempt to keep
$d_1$ and $d_2$ as given by vector meson resonance
saturation and then fit $d_3$ to the experimental value of the $\eta$--decay
width \cite{Ko95} leads to a very large $d_3$ and implies that further
terms in the Lagrangian are required for a realistic description of pion
polarizabilities and the the process
$\gamma \gamma \to \pi^0 \pi^0$. In summary, one is led to the conclusion
\cite{Belk95} that a self--consistent, quantitative description of
the decay $\eta \to \pi^0 \gamma \gamma$ based on the Lagrangians
(\ref{Bijlag}) and (\ref{njllag}) is problematic.

The numerical decay widths quoted so far are calculated from pure
${\cal{L}}_6$ tree--level amplitudes.
A complete treatment of the ${\cal{L}}_2$ two--loop contributions has been
achieved in an SU(2)--calculation
of $\gamma \gamma \to \pi^0 \pi^0$
\cite{Gas94}, where only pionic degrees of freedom are taken into account.
Numerically, the two--loop amplitude is crucial for a fit of the experimental
spectrum. For the $\eta$--decay, the contributions
of the factorizable 2--loop diagrams
have been given recently \cite{Belk95}. Although the effects
on the $\eta$--decay are small,
it is clearly desirable to carry out a complete calculation
of the two neutral meson processes at two--loop level.

The purpose of the present paper is thus twofold: First, we explore the
structure of the tree--level amplitudes if the most general Lagrangian
${\cal{L}}_6$ is used and compare the result to previous calculations.
Second, we outline a method to derive the two--loop
amplitude including overlapping graphs. We present the results for the case
of pure pion loops and estimate the missing parts for
$ \eta \to \pi^0 \gamma \gamma $. A method of solving the specific
Feynman integrals occurring in the problem may be of general interest
and is therefore discussed in the appendix.

\section{The $\eta \to \pi^0 \gamma \gamma$ Amplitude}

\subsection{General Formalism and Kinematics \label{Notation}}

Throughout the following, we will use the notation of \cite{Sche94} applied
to our special case: The $U(3)$--matrix $U=\exp({i \over F_{\pi}}
(\Phi_8 + \Phi_1))$  contains the nonet of pseudoscalar mesons
\begin{equation}  \Phi_8(x)=\left(\matrix{ \pi^0 + {\eta_8 \over \sqrt{3}} &
\sqrt{2} \pi^+ &
\sqrt{2} K^+ \cr \sqrt{2}  \pi^- & - \pi^0 + {\eta_8 \over \sqrt{3}} &
\sqrt{2} K^0
\cr \sqrt{2} K^- & \sqrt{2} \bar K^0 &  {-2 \eta_8 \over \sqrt{3}} \cr
} \right); \qquad \Phi_1(x)=  {\sqrt{ 2  \over 3} } \eta_1 I \label{mixing}
\end{equation}
$F_{\pi}=93.2$ MeV is the pion decay constant \cite{Gas85}.
The covariant
derivative for photons as external gauge fields reads
$$ D_{\mu} B=\delta_{\mu} B+ i e A_{\mu} [Q,B] $$
where $Q={\rm diag}(2/3, -1/3, -1/3)$ is the quark charge matrix
and $B$ stands for be any operator that transforms linearly under the
chiral group $U(3)_L \times U(3)_R$: $B \to B'=V_R B V_L^\dagger $
We note that
\begin{equation} [Q,\Phi] =
\left(\matrix{ 0  & \pi^+ & K^+ \cr \pi^- & 0 & 0
\cr K^- & 0 & 0 \cr
} \right)  \label{qkomm} \end{equation}
so that the commutator $[Q,\Phi]$ couples the photon to charged mesons only.
In Eq. (\ref{Bijlag}), coupling of the neutral mesons to photons is mediated
by the field tensor $F_{\mu \nu} =( \delta_{\mu} A_{\nu} - \delta_{\nu}
A_{\mu}) $.

The physical $\eta$ particle can be written as \cite{Bij92}
\begin{equation} \eta = \cos \theta \eta_8 - \sin \theta \eta_1 \sim {2 \over
3}
\sqrt{2}  \eta_8 + {1 \over 3} \eta_1 .\label{etaphys} \end{equation}
In our normalization, the lowest order chiral Lagrangian reads
\begin{equation}  {\cal{L}}_2={F_{\pi}^2 \over 4} Tr(D_{\mu} U D^{\mu}
U^{\dagger})
+    {F_{\pi}^2 \over 4} Tr(\chi^{\dagger} U + U^{\dagger} \chi)
\label{L2} \end{equation}
where the mass terms contain the quark mass matrix
$\chi=2 B_0 {\rm diag}(m_u, m_d, m_s)$ and the constant $B_0$
 relates the masses
of quarks and pseudoscalar mesons via \cite{Gas85,Bij92}
\begin{equation}  B_0 = {m_K^2 \over m_u + m_s }
= {m_{\pi}^2 \over m_u + m_d } = {\Delta m_K^2 \over m_d - m_u }.
\end{equation}
A numerical estimate for the electromagnetic mass split
of the kaon yields the value of $\Delta m_K \sim 6000$ MeV$^2 $ \cite{Bij92}.

The general form of the amplitude for the decay $\eta(P) \to \pi^0(p)
\gamma(q_1, \epsilon_1) \gamma(q_2, \epsilon_2)$ is \cite{Pic92,Moz89}
\begin{eqnarray} M& = &
 \epsilon_{1 \mu} \left\{ A(s,t) \cdot \left( g^{\mu \nu} {s \over 2}
- q_2^{\mu}
q_1^{\nu} \right) \right. \cr
\mbox{} & & \left. + B(s,t) {2 \over s}  \cdot \left( g^{\mu \nu} P\cdot q_1
P\cdot q_2 + P^{\mu} P^{\nu} {s \over 2} - q_2^{\mu} P^{\nu} P \cdot q_1
- P^{\mu} q_1^{\nu} P \cdot q_2 \right) \right\} \epsilon_{2, \nu}
\label{eich} \end{eqnarray}
where we have defined the kinematic invariants $s = (q_1+q_2)^2$,
$t = (P-q_2)^2$,
$u = (P-q_1)^2 = m_{\eta}^2 +m_{\pi}^2 - s -t$.
The decay rate is calculated according to
\begin{equation}  {d \Gamma \over ds}= {1 \over 1024 m_{\eta}^3 \pi^3 }
\int_{t_1}^{t_2}
dt \left( \mid A s - m_{\eta}^2 B \mid^2 + { \mid B \mid^2 \over s^2}
(m_{\pi}^2 m_{\eta}^2-tu)^2 \right), \label{decayrate} \end{equation}
where s is restricted to $0 \leq s \leq (m_{\eta}^2 -m_{\pi})^2$ and
\begin{equation}   t_{1,2} = {1 \over 2} \left[ (m_{\eta}^2+m_{\pi}^2-s)
\pm \sqrt{
   (m_{\eta}^2+m_{\pi}^2-s)^2-4 m_{\eta}^2 m_{\pi}^2} \right]. \end{equation}

\subsection{Complete ${\cal{L}}^6$--Amplitude}

The tree--level amplitude from the Lagrangian ${\cal{L}}^6$ forms the
dominant contribution of the $\eta$--decay width and will therefore be
discussed separately from the one-- and two--loop corrections.

The most general chiral Lagrangian at ${\cal{O}}(p)^6$ has been derived
for the $SU(3)$--case $U=\exp ( {i \over F_{\pi}} \Phi_8)$.
Introduction of $\eta - \eta'$ mixing according to Eq. (\ref{mixing})
eliminates a symmetry constraint and
potentially generates more terms that we have to neglect at this point
(for an example see \cite{Bij88a}).
Keeping this simplification in mind, we can take the ${\cal{L}}^6$--terms
required from Table II, section
no. 4 (terms involving $B_{30}-B_{50}$) of Ref. \cite{Sche94}.
The number of independent
structures contributing to the 2 photon/2 neutral meson processes
can be determined in two steps: First make use of the simplifications
explained in Sections IV B/C of Ref. \cite{Sche94} and note that, owing to
Eq. (\ref{qkomm}), the commutator $[Q,\Phi]$ can be set to zero in the
covariant derivative as well as under the trace of the
${\cal{L}}^6$--expressions. As a result, for instance
both the terms in $B_{35}$ and
$B_{36}$ of Ref. \cite{Sche94}, Table II reduce to the $d_2$--term in
Eq. (\ref{Bijlag}):
\begin{eqnarray} & B_{35} Tr([D_{\alpha}U]_- [D^{\beta}U]_- [G^{\alpha
\gamma}]_+
[G_{\beta \gamma}]_+ ) +
B_{36} Tr([D_{\alpha}U]_- [D^{\beta}U]_- [G_{\beta \gamma}]_+
[G^{\alpha \gamma}]_+ )& \cr \mbox{} & =
d_2 F_{\mu \alpha} F^{\mu \beta} Tr(Q^2 \delta^{\alpha}
 U \delta_{\beta} U^{\dagger}), \quad d_2=4(B_{35}+B_{36}).&  \end{eqnarray}
The terms resulting from this procedure are more symmetric
than the original general ${\cal{L}}^6$--terms.
This implies that terms which are independent in the general Lagrangian
${\cal{L}}^6$ can be redundant due to $[Q,\Phi]=0$.
Therefore, in a second step,
we have to check the result of the simplified Lagrangian and reject
terms that are linked by a trace relation of the
type discussed in \cite{Sche94}, App.\ A. For example, one of the
terms
\begin{eqnarray}  B_{34} Tr([D_{\mu}U]_- [D^{\mu}U]_- [G^{\alpha \beta}]_+
[G_{\alpha \beta}]_+ ) & = & 4 B_{34}  F_{\alpha \beta} F^{\alpha \beta}
Tr(Q^2 \delta^{\mu}  U \delta_{\mu} U^{\dagger}),  \cr
   B_{37} Tr([D_{\mu}U]_- [D^{\mu}U]_-) Tr( [G^{\alpha \beta}]_+
[G_{\alpha \beta}]_+ ) & = & 4 B_{37}  F_{\alpha \beta} F^{\alpha \beta}
Tr(Q^2) Tr( \delta^{\mu}  U \delta_{\mu} U^{\dagger}),  \cr
   B_{38} Tr([D_{\mu}U]_- [G^{\alpha \beta}]_+) Tr([D^{\mu}U]_-
[G_{\alpha \beta}]_+ ) & = & 4 B_{38}  F_{\alpha \beta} F^{\alpha \beta}
Tr(Q \delta^{\mu} U ) Tr(Q  \delta_{\mu} U^{\dagger})  \end{eqnarray}
is redundant after simplification due to the the trace relation
$$ 4 Tr(A^2 B^2) + 2 Tr(ABAB) - Tr(A^2)Tr(B^2)-2 (Tr(AB))^2 =0 $$
for any complex $3 \times 3$--matrices $A$, $B$ with $Tr(A)=Tr(B)=0$.
Alternatively, the $B_{34}$ and $B_{37}$ can be seen to be equivalent
because of $Tr(Q^2)=2/3$.

We are finally left with the 3 single trace terms of Eqs. (\ref{Bijlag})
and (\ref{njllag}) plus 3 double trace terms
\begin{eqnarray}
d_1 F_{\mu \nu} F^{\mu \nu} Tr(Q^2 \delta^{\lambda}
U \delta_{\lambda} U^{\dagger})
 & + & d_4  F_{\mu \nu} F^{\mu \nu}
Tr(Q \delta^{\mu}  U ) Tr( Q \delta_{\mu} U^{\dagger})
 \cr \mbox{}  +
d_2 F_{\mu \alpha} F^{\mu \beta} Tr(Q^2 \delta^{\alpha}
 U \delta_{\beta} U^{\dagger})
 & + &   d_5 F_{\mu \alpha } F^{\nu \beta}
Tr(Q \delta^{\alpha}  U) Tr(Q \delta_{\beta} U^{\dagger})
 \cr \mbox{}  +
d_3 F_{\mu \nu} F^{\mu \nu}
Tr(Q^2 (\chi^{\dagger} U + U^{\dagger} \chi))
 & + & d_6  F_{\mu \nu} F^{\mu \nu}
Tr(Q^2) Tr(\chi^{\dagger} U + U^{\dagger} \chi)).  \label{twotracelag}
 \end{eqnarray}
and thus a total of 6 low--energy constants to be determined either from
experiment or in some model\footnote{Elimination of redundant terms
in the Lagrangian is desirable for simplicity but not indispensable from
a practical point of view. Keeping more terms would result in fixed
linear combinations of coefficients appearing in the amplitudes; the
fit procedure would ultimately not be affected.
Moreover, we have some freedom as how to represent the Lagrangian and which
terms to discard. A choice different from Eq. (\ref{twotracelag}) might affect
the expressions for the amplitudes and the values of the parameters quoted
but of course must leave the physical content of the Lagrangian invariant. }.
The low--energy constants $d_1, \ldots , d_6$ are related to the $B$--constants
of \cite{Sche94} as follows:
\begin{eqnarray}
d_1 & = & 4 B_{34} + {8 \over 3} B_{37}, \; \,
 \qquad  \qquad  d_4 = 4 B_{38}, \cr
d_2 & = & 4 (B_{35}+B_{36})+ {8 \over 3} B_{39}, \quad
d_5 = 4 (B_{40}+ B_{41}), \cr
d_3 & = & 4 B_{47}, \qquad \qquad \qquad \qquad d_6 = 4 B_{50}.
\label{bconstants} \end{eqnarray}

Using Eq. (\ref{etaphys}),
the $\eta$--decay amplitude from the full ${\cal{L}}^6$--Lagrangian can
now be cast in the form (\ref{eich}) with
\begin{eqnarray} A_{\eta}(s,t)&=& {2 \sqrt{2} \over 3 \sqrt{3} F_{\pi}^2}
\left\{ (4 d_1 - 12 d_4) m_{\eta}^2 - 2 d_3 m_{\pi}^2 - (4 d_1 - 12 d_4
+ d_2 - 3 d_5) P \cdot (q_1+q_2) \right\} \cr
B_{\eta} (s,t)&= & {4 \sqrt{2} \over 3 \sqrt{3} F_{\pi}^2} {s \over 2}
\left\{ d_2 - 3 d_5 \right\} \label{etaampli}
\end{eqnarray}
Note that we used the isospin approximation $m_u=m_d$ neglecting
a small contribution from the double trace mass term in Eq. (\ref{twotracelag})
proportional to $d_6$.  This does not change any of the conclusions of this
chapter. The $d_6$--terms does
affect, however, the $\gamma \gamma \to \pi^0 \pi^0$ amplitude
which can be written in analogy to Eq. (\ref{etaampli})
\begin{eqnarray} A_{\pi}(s,t)&=& -{4 \over F_{\pi}^2} (s-2 m_{\pi}^2)
 ({5 \over 9} d_1 - d_4) -
{1 \over F_{\pi}^2} s ({5 \over 9} d_2 - d_5) -
{4 \over F_{\pi}^2} m_{\pi}^2 ({5 \over 9} d_3 + {12 \over 9} d_6)
\cr
B_{\pi} (s,t)&= &
{4 \over F_{\pi}^2} {s \over 2}
 ({5 \over 9} d_2 - d_5). \label{piampli}
\end{eqnarray}
Determination of the 6 constants $d_1 \ldots d_6$ from experimental data
alone is now impeded by a combination of factors:

1) For the process $\eta \to \pi^0 \gamma \gamma$ where the
${\cal{L}}^6$--amplitude (\ref{etaampli}) forms the leading contribution,
only the total decay
width  has been measured \cite{Data90}.

2) The energy dependence of the cross section
for neutral pion pair production is experimentally known. In this case,
however, the amplitude (\ref{piampli}) is small and interferes with the
large, complex one--loop and two--loop
contributions \cite{Gas94} so that the fit becomes ambiguous.
Complementary data such as pion
polarizabilities \cite{Gas94,Anti83} don't put sufficient constraints on
the parameters.

3) There are further two neutral meson/two photon data, e.\ g.\ the production
cross sections for $\gamma \gamma \to
\eta \pi^0$ and $\gamma \to  K^0 \bar K^0$ \cite{Data94}, but the processes
have too high energy thresholds.

On the other hand, the method of resonance saturation has been successfully
applied in order to confirm the phenomenological constants of the
Gasser--Leutwyler Lagrangian ${\cal{L}}^4$ \cite{Eck89}.
Moreover, the pion production amplitude (\ref{piampli}) with the
constraint $d_4=d_5=d_6=0$ has been determined via resonance saturation
\cite{Gas94} and compares well with the data.
We proceed to discuss how the method can be applied to our general case.

\subsection{Meson Resonance Amplitude \label{VMDmod}}

In the VMD model, the leading contribution to $\eta \to \pi^0 \gamma \gamma$
is generated by exchange of internal vector mesons $\omega$, $\rho$. This
generates the amplitudes \cite{Pic92,Bij92,Gas94}:
\begin{eqnarray} A_{\eta}^{VMD}(s,t)& = & - \sum_{V=\omega,\rho, \Phi}
{ G_{\eta,  V} \over 2} \left[
{t + m_{\eta}^2 \over t-m_V^2 } + {u + m_{\eta}^2 \over u-m_V^2 } \right]
\to   \sum_{V=\omega,\rho, \Phi} {G_{\eta V} \over m_V^2}
 \left(3 m_{\eta}^2+m_{\pi}^2-s \right), \cr
 B_{\eta}^{VMD}(s,t)& = & - \sum_{V=\omega,\rho} G_{\eta V} {s \over 2} \left[
{1 \over t-m_V^2 } + {1 \over u-m_V^2 } \right]
\qquad \qquad \to   \sum_{V=\omega,\rho} {G_{\eta V}\cdot s \over  m_V^2}
\label{etavmd} \end{eqnarray}
The right hand side indicates the ${\cal{O}}(p^6)$ low energy limits of the
VMD amplitudes. The VMD amplitude for neutral pion pair
production $\gamma \gamma \to \pi^0 \pi^0$ is (\ref{etavmd})
with the substitutions $G_{\eta V} \to G_{\pi V}$, $m_{\eta} \to m_{\pi}$.
The coupling constants $G_{\eta V}$, $G_{\pi V}$, can be extracted from
the decay widths of the vector mesons. We list the experimental values
\cite{Ng92,Gas94} of all the coupling constants used in the normalization
of Eqs. (\ref{etavmd})--(\ref{etascal})
together with the meson masses in Table \ref{VMDcon}.

The contributions of the C-odd axial-vector resonances are \cite{Ko93}
\begin{eqnarray} A_{\eta}^{B}(s,t)& = & - \sum_{B=b_1,h,h'}
{ G_{\eta,  B} \over 2} \left[
{-t + m_{\eta}^2 \over t-m_V^2 } + {-u + m_{\eta}^2 \over u-m_V^2 } \right]
\to   \sum_{B=b_1,h,h'} {G_{\eta B} \over m_V^2}
 \left(3 m_{\eta}^2+m_{\pi}^2-s \right), \cr
 B_{\eta}^{B}(s,t)& = & - \sum_{B=b_1,h,h'} G_{\eta V} {s \over 2} \left[
{1 \over t-m_V^2 } + {1 \over u-m_V^2 } \right]
\qquad \qquad \to   \sum_{B=b_1,h,h'} {G_{\eta V}\cdot s \over  m_V^2}
\label{etaaxi} \end{eqnarray}
Again, there is an equivalent amplitude for pion pair production.
The axial vector mesons interfere constructively with the vector meson
amplitudes (\ref{etavmd}) and thus enhance the $\eta$--decay width
while deteriorating the reproduction of the pion pair production data
\cite{Ko93}. We include only the measured $b_1$ resonance in our model.

The production cross section $\gamma \gamma \to \eta \pi^0$ \cite{Data94}
is dominated by the (scalar) resonance $a_0(983)$ and the (tensor)
resonance $a_2(1318)$. Their contributions can be written as
\begin{equation} A_{\eta}^{T}(s,t) =  -
{ G_{\eta}^{T} \over 4} {m_{\eta}^2  \over m_T^2-s }
; \qquad
 B_{\eta}^{T}(s,t) =
{ G_{\eta,  T} \over 2} {s \over 2} {1 \over m_T^2-s }
\label{etatens} \end{equation}
for the $a_2$ and
\begin{equation} A_{\eta}^{S}(s,t) =  {
\left( G_{\eta}^{Sd} (s+m_{\eta}^2-3 m_{\pi}^2)
 + G_{\pi}^{Sm} 2 m_{\pi}^2 \right) \over m_S^2-s}
; \quad  B_{\pi}^{S}(s,t) =  0;
\quad  G_{\pi}^{Sd} \cdot G_{\pi}^{Sm} \geq 0
\label{etascal} \end{equation}
for the $a_0$ respectively \cite{Eck89,Data90}.
The $\gamma \gamma \to \pi^0 \pi^0$ spectrum shows the
scalar resonance $f_0(983)$ and the tensor resonance $f_2(1275)$ with
contributions analogous to Eq. (\ref{etatens}), (\ref{etascal})
(set $m_{\eta} \to m_{\pi}$ and choose the appropriate
couplings, see \cite{Gas94}).

The coupling constants for the vector and axial--vector mesons are positive,
but the signs of the (s-channel and therefore not quadratic)
coupling constants $G^{S}$ and $G^{T}$ relative to the
VMD amplitude are ambiguous. Our choice is motivated by the following
observation: Consider the one-loop and (approximate)
analytic two--loop amplitude for $\gamma \gamma \to \pi^0 \pi^0$ from
Ref. \cite{Gas94} as the QCD--background for the process and complement
this amplitude by the {\it full resonance amplitude} of Eqs.
(\ref{etavmd}), (\ref{etaaxi}), (\ref{etatens}) and (\ref{etascal}).
At high energy, parametrize
the resonance widths $m_T$ and $m_S$ by a relativistic Breit--Wigner
form given by Eqs. (5) and (6) of the original
data analysis of the Crystal Ball experiment \cite{Mar90}. The result is
of course not accurate in the region of the $f_0$ and $f_2$ resonances but fits
the {\it shape} of the data fairly well provided the signs of
the couplings are chosen as in Table I (see Fig. \ref{pivmd}).
With the opposite signs, the measured cross section can not be reproduced.
Our sign convention implies that the $\eta$--decay width is {\it increased}
by the contribution of the scalar and tensor mesons.
By the way, this result is consistent with the interference pattern found
for the scalar resonances in the extended Nambu--Jona-Lasinio model of
\cite{Belu95}.

Having constructed a realistic meson exchange model for the two neutral
meson/two photon processes (see Fig. \ref{pivmd}), we can now proceed to
test the prediction of the chiral Lagrangian ${\cal{L}}^6$ to both pion
pair production and the $\eta$--decay width. To this end, the propagators
of Eqs. (\ref{etavmd}) -- (\ref{etascal}) are taken to first order in
the kinematic invariants $s, \, u$ and $t$ as indicated in Eqs. (\ref{etavmd})
and (\ref{etaaxi}); the resulting ${\cal{O}}(p^6)$ low--energy
expressions are used to fit
simultaneously the parameters of the chiral amplitudes Eqs. (\ref{etaampli})
and (\ref{piampli}). The procedure yields a total $\eta$--decay width
of
\begin{equation}
 \Gamma^{L^6} (\eta \to \pi^0 \gamma \gamma) = 0.652 \, {\rm eV}.
\label{treeamp} \end{equation}
(The last digit has been quoted for the discussion of the small
loop contributions, see below).
As shown in Fig. \ref{etafit}, the spectrum obtained with the
${\cal{O}}(p^6)$--fit differs considerably from the original meson exchange
spectrum, being smaller as $s \to 0$ and showing a broad maximum at
$\sqrt{s}=E_{\gamma \gamma} \sim {3 \over 4} \sqrt{s_{max}}$.
The difference at low
invariant mass $\sqrt{s}$ is induced by the higher order terms in the
t--channel amplitudes Eqs. (\ref{etavmd}) and (\ref{etaaxi}). As $s$
grows, we increasingly neglect contributions from the s--channel amplitudes
(\ref{etatens}) and (\ref{etascal}) as well. It is in fact easy to perform
an "all order" fit of the meson--exchange amplitude, see Fig. \ref{etafit}.
We discard this option, however, as not consistent with the spirit of the
momentum expansion.

The ${\cal{O}}(p^6)$ tree--level contributions are small for low--energy
pion pair production so that details of the fit have only a minor effect
on the cross section (see Fig. \ref{pifit}). $\gamma \gamma \to \pi^0 \pi^0$
therefore doesn't put further restrictions on the parameters of the model.
Note, however, that it is
in general not possible to obtain an accurate reproduction of the (all order)
meson exchange amplitude by ${\cal{L}}^6$--terms in the energy domain
$E_{\gamma \gamma} > 500$ MeV.

The  coupling constants $d_1 \ldots d_6$ derived from the fit and collected
in Table \ref{l6const} deserve some comments. At first glance, a comparison
of the ${\cal{O}}(p^6)$ and all order results for the kinetic constants
$d_1, \, d_2, \, d_4$ and $d_5$
suggests that the fit procedure is highly unstable, changing the order of
magnitude and even the signs of the constants as one passes from one
scheme to an other. This behavior is indicative for the fact
that the ratio between the empirical
meson coupling constants $G_{\eta}$ and $G_{\pi}$
used in the fit (Table \ref{VMDcon}) is close to the ratio between the
SU(3)--factors contained in Eqs. (\ref{etaampli}) and (\ref{piampli}) --
quark symmetry and therefore the physics of the process itself seems to
exclude a stable inversion procedure. The results
appear more trustworthy if one keeps in mind that in both amplitudes,
only linear combinations of the coefficients $d_1 \ldots d_6$ enter. For
the $\eta$--decay (Eq. (\ref{etaampli})), these are

a) $d_1-3d_4=4 B_{34} + {8 \over 3} B_{37} - 12 B_{38} =
-3.56 \cdot 10^{-3}$ GeV$^{-2}$
($-4.81 \cdot 10^{-3}$ GeV$^{-2}$) for the ${\cal{O}}(p^6)$ (all order) fit,
to be compared with the value of $-3.6 \cdot 10^{-3}$ GeV$^{-2}$ derived
in the simplest fit scheme (Eq. (\ref{Bijconst})).

b) $d_2-3d_5=4 (B_{35} + B_{36})   + {8 \over 3} B_{39} - 12 (B_{40}+ B_{41})
 = 9.59 \cdot 10^{-3}$ GeV$^{-2}$
($12.4 \cdot 10^{-3}$ GeV$^{-2}$) for the ${\cal{O}}(p^6)$ (all order) fit,
to be compared with the value of $7.2 \cdot 10^{-3}$ GeV$^{-2}$
from Eq. (\ref{Bijconst}).

The linear combinations ${5 \over 9} d_1 - d_4$ and
${5 \over 9} d_2 - d_5$ that appear the pion production amplitude
are similarly stable with respect to the fit method chosen. Due to the
similarity of our meson exchange model with the one used  in \cite{Gas94},
they are doomed to be consistent with the constants found there.

The mass term constants $d_3$ and $d_6$ show a slightly different behavior.
As $d_6$ doesn't influence the $\eta$--decay amplitude, they have to be
stable {\it separately}, which is indeed what we observe. The values found
for $d_3$ are consistent with the value fitted to the experimental total
decay width \cite{Ko95} ($d_3=45 \cdot 10^{-3}$ GeV$^{-2}$) within $\sim 10\%$,
furthermore, \cite{Ko95} obtains a spectrum similar to the one shown in
Fig. \ref{etafit}.
Note, however, that we have derived our coupling constants from the
general form of the Lagrangian ${\cal{L}}^6$ and the meson exchange model
alone:
no assumption concerning the $\eta$--decay width or pion production cross
section was made. The agreement between our model and the one discussed in
\cite{Ko95} might indicate that the relatively high $\eta$--decay width
measured is indeed correct.

The mass term difference $\Delta_m =
{5 \over 9} d_3 + {12 \over 9} d_6 = {4 \over 9} (5 B_{47} + 12 B_{50})=
1.44 \cdot
10^{-3}$ GeV$^{-2}$ entering the pion production amplitude (\ref{piampli})
is more than an order of magnitude below the value of $d_3$ (and
thus of the order of magnitude of the kinetic constants). Again,
our values of $\Delta_m$ have to be consistent with the fit of
\cite{Gas94} by construction. However, identifying $\Delta_m$ with the
mass term $d_3$ entering the $\eta$--decay amplitude would inevitably lead
to an underprediction of the $\eta$--decay width.

It seems therefore that any attempt to describe $\eta \to \pi^0
\gamma \gamma$ and $\gamma \gamma \to  \pi^0 \pi^0$ simultaneously in a model
containing
only 3 ${\cal{L}}^6$ constants as in Eqs. (\ref{Bijlag}) and (\ref{njllag})
is bound to be troubled by the incompatibility of the mass terms.
This explains some of the inconsistencies about the constants $d_1$, $d_2$
and $d_3$ found in the literature.

Because of to the relative
insensitivity of the $\gamma \gamma \to \pi^0 \pi^0$ cross section, a
measurement of the $\eta$--decay spectrum would discriminate
between different sets of ${\cal{L}}^6$ coupling constants. However, in
order to pin things further down, we need an accurate estimate of the
loop contributions generated in
chiral perturbation theory.

\subsection{Effect of Chiral Loops to ${\cal{O}}(p^6)$ }

The following section is devoted to a discussion of the size of the
various loop contributions occurring up to ${\cal{O}}(p^6)$ and the basic
technicalities of their evaluation.
For the formal aspects of an two--loop calculation and the
implications for the amplitude, we refer to earlier publications
\cite{Gas94,Golo95}.

The leading loop contributions appearing in chiral perturbation theory are
the ${\cal{O}}(p^4)$ one--loop diagrams with vertices generated by the
Lagrangian ${\cal{L}}^2$ of Eq. (\ref{L2}). A detailed calculation of this
contribution for the $\gamma \gamma \to \pi^0 \pi^0$ amplitude has
been given in \cite{Hol88}; the $\eta$--decay can be treated analogously,
employing expressions similar to Eq. (\ref{sixpi}) without the tadpole
term. For the result see Ref. \cite{Bij92}\footnote{From the $\eta \pi^0 \pi^+
\pi^-$--vertex displayed below (Eq. (\ref{vertex})), we obtain only
the first (leading) term of the one pion loop amplitude Eq. (14) of Ref.
\cite{Bij92}. We agree, however, with the numerical result Eq. (16) of
this reference.}. The ${\cal{O}}(p^4)$--amplitude
involves charged pion and kaon loops and strongly interferes with the
tree--level amplitude from ${\cal{L}}^6$ so that its effect increases
with the latter amplitude. Adding it to the ${\cal{L}}^6$ result
Eq. (\ref{treeamp}), the decay width increases by about 10$\%$ to
\begin{equation}
 \Gamma^{L^6+O(p^4)} (\eta \to \pi^0 \gamma \gamma) = 0.733 \, {\rm eV}.
\label{width1loop} \end{equation}
We note that about 3/4 of the one--loop effect stems from charged kaon loops.
Inclusion of pion loops alone would increase the decay width to 0.668 eV.

At order ${\cal{O}}(p^6)$, there are further one--loop diagrams generated
by the Gasser--Leutwyler Lagrangian ${\cal{L}}^4$ \cite{Gas85}.
Besides the recalculation of vertices for all
the ${\cal{L}}^4$--terms, these contributions introduce no new technical
difficulties. For a pion loop only, the task is simplified because, like
in the case of the Lagrangian ${\cal{L}}^2$, $\eta \pi^0 \pi^+ \pi^-$--vertices
can only be generated by the mass terms. This leaves the terms in $L_4$
and $L_5$ as possible candidates where the renormalized coupling constant
$L^r_4$ is $N_c$--suppressed and empirically consistent with 0.
The $L_5$--amplitude invokes the loop function $F$ defined in the Appendix and
is compact enough to be displayed
\begin{eqnarray}
A_{L_5}^{\pi+ \pi^-}(s)
& = & - L_5^r {8 \sqrt{2} \over 3 \sqrt{3}} {e^2 (\Delta m_K)^2
\over (4 \pi)^2 F_{\pi}^4} \left( 24 s- 36 m_{\pi}^2)-4 m_{\eta}^2 \right)
  {1 \over s} \left( {1 \over 2} + {m^2 \over s} F(s) \right) \cr
B_{L_5}^{\pi+ \pi^-}(s) & = & 0; \qquad L_5^r=2.2\times 10^{-3}.
\label{l4amp} \end{eqnarray}
The decay width from the amplitude (\ref{l4amp}) turns out to be very
small ($\Gamma(L^4, \pi^+ \pi^-) = 0.45 \times 10^{-4}$ eV).
In superposition to the amplitude
corresponding to the decay width (\ref{width1loop}), its effect is totally
negligible.

The calculation of charged kaon loops from ${\cal{L}}^4$ is analogous to
the pion loop calculation but invokes the terms in the Gasser--Leutwyler
Lagrangian proportional to $L_4$, $L_5$, $L_8$ and $L_{10}$. By interference
of all terms, we gain a factor of 25
($\Gamma(L^4, \pi^+ \pi^- + k^+ K^- ) = 1.9 \times 10^{-3}$ eV), the complete
${\cal{L}}^4$--contribution reduces the decay width (\ref{width1loop})
to
\begin{equation}
 \Gamma^{L^6+O(p^4)+L^4} (\eta \to \pi^0 \gamma \gamma) = 0.673 \, {\rm eV}.
\label{widthl4loop} \end{equation}
The values quoted for the ${\cal{L}}^4$--loops seem to match the
$Z_A^2=0.62$ results of Ref. \cite{Belk95}.
We come to the conclusion that the one--loop contributions from
${\cal{L}}^2$ and ${\cal{L}}^4$ mostly cancel each other.

The next group of diagrams are the factorizable 2--loop diagrams derived
from Fig. \ref{Configurations} (a) and (b). Again, we first consider the
case where only pions are propagating in the loops. If a loop has no photons
attached to it, there are also neutral pions allowed. Thus, the
6--meson vertex diagram of Fig. \ref{Configurations} (a) contains two
charged pion loops plus the combination of one charged and one neutral
loop.

It is a straightforward procedure to determine the Feynman diagrams and
derive the corresponding integral expressions. Solving these integrals is
largely simplified by the fact that they effectively reduce to a product
of two one--loop integrals. On the other hand,
as the masses of external and loop particles
are comparable, we need to solve the integrals exactly.
This can be done by putting
propagators with different loop momenta together by means of Feynman parameters
and performing the four dimensional integration in the scheme of dimensional
regularization \cite{Thooft72}. One is left with an (at most) two--dimensional
integration in the Feynman parameters space that is analytic due to the
fact that the external photons are on their mass shell. It is
further simplified because the masses
of the loop particles are all equal. The task is nevertheless a bit tedious
because the one--loop integrals generate divergent terms
proportional to $\lim_{n \to 4} {2 \over 4-n}$ if the number $n$ of
space--time dimensions is taken to 4  so that one is forced to
expand all the expressions to {\it linear} order in $\epsilon={4-n \over 2}$.
Details of the calculation are discussed in the Appendix, subsection B.

It is convenient to express everything as sums of
products of the elementary integrals $I_{tad}$, $I_{bub}$ etc.
(Eqs. (\ref{tadpole}), (\ref{bubble}), (\ref{triangle})) and then do the
multiplication of one--loop integrals
and subtraction of singularities numerically. In order to
be consistent with the scheme of chiral perturbation theory and the
renormalization of the ${\cal{L}}^4$ coupling constants, we have to
subtract terms proportional to powers of
$R \equiv -{1 \over \epsilon} -1 - \ln(4 \pi)
+ \gamma$, where $\gamma$ is the Euler constant. Also, the renormalization
constant is to be chosen as $\mu = m_{\eta}$ in order to be consistent with the
numerical values quoted for the ${\cal{L}}^4$ low--energy constants.
The finite end result has to be gauge invariant, i.\ e.\ proportional to
$g^{\mu \nu} {s \over 2} - q_2^{\mu} q_1^{\nu}$. (As in the case of the
one--loop amplitudes, there is no contribution $B(s,t)$.)
Gauge invariance can be shown analytically and, on the other hand,
serves as a check to the calculation.

It turns out that the various terms in the factorizable two pion--loop
amplitude largely cancel each other.
Our result for the decay width is $\Gamma = 2.7 \times 10^{-4}$ eV,
about a factor of 3 smaller than the
one pion--loop amplitude from ${\cal{L}}^2$ and
well in agreement with the result of \cite{Belk95}.
The effect is also very small in superposition
\begin{equation}
 \Gamma^{L^6+O(p^4)+L^4+{\rm fact.}} (\eta \to \pi^0 \gamma \gamma) = 0.682
 \, {\rm eV}.
\label{widthfacloop} \end{equation}

Due to the crucial role of cancellations between various terms in the
amplitude, it is problematic to predict contributions without actually
performing the calculation. We may, however, observe the following:
Replacement of a pion loop by a kaon loop lifts the G--parity suppression
expressed by the factor $(\Delta m_K)^2/m_K^2$ but introduces a mass factor
from
the propagators which, in the low--energy limit, amounts to $m_{\pi}^2/m_K^2$.
A very rough estimate would therefore be that the amplitude from a kaon loop
is  larger than the corresponding pion loop amplitude by a factor of
$m_{\pi}^2 / (\Delta m_K)^2 \sim 3$, a factorizable two--loop diagram
containing two kaon loops should be suppressed by $m_{\pi}^4 / (\Delta m_K)^2
m_K^2 \sim 1/3$.
The values 3 and 1/3 quoted are the relevant ratios because only the
interference terms between loop and tree--level contributions affect the
result for the decay width noticeably.
Our estimate about correct for the ${\cal{L}}^2$ one--loop contributions
and seems to be confirmed by the two--loop results quoted in \cite{Belk95}.
(Estimating the kaon loops from ${\cal{L}}^4$ is more problematic because
more structures in the Lagrangian can generate kaon loops.) Applying it
to the factorizable two--loop amplitude, we would conclude that the
two--kaon loops should be totally negligible whereas the pion--kaon loop
amplitude might change the result (\ref{widthfacloop}) by $\pm 0.02$ eV,
a value that is markedly lower than the uncertainties introduced by the
meson exchange model.

The technically challenging part of the amplitude consists in evaluating
the overlapping two--loop diagrams Fig. \ref{overlapdiag}.
Again, we need the {\it exact} result and can not resort, e.\ g.\ to
developments in the external momenta that have been successfully applied
for QCD calculations \cite{Hoog85}. Feynman diagrams of the type
required have been evaluated for the $\gamma \gamma \to
\pi^0 \pi^0$ amplitude. As outlined in Ref. \cite{Gas94}, the expressions
corresponding to Fig. \ref{overlapdiag} (a) and (b) can indeed be transformed
into two--dimensional integrals by first integrating over one of the
loop momenta and then transforming the result into a dispersion integral over
a parameter--dependent box diagram that in turn can be represented as a
dispersion integral. We note that the graph \ref{overlapdiag} (a)
leads to direct $s-t$ channel box diagrams suggesting a fixed--$t$ dispersion
relation whereas the "master"--diagram \ref{overlapdiag} (b) yields
crossed $u-t$ channel box diagrams. In our case, the analyticity properties
of the graphs are unfortunately more complex because the external
$\eta$ is unstable against decay and thus introduces anomalous thresholds
\cite{Itz86}.
Correspondingly, we will find that the overlapping amplitude has a complex
part for arbitrary provided that $m_{\eta} \geq	3 m_{\pi}$.

Our solution avoids dispersion relations and is inspired by
treatments such as \cite{Velt84} where the graphs are transformed into
expressions of the type $I(m,n,Z)$ (see Eq. (\ref{hierarchy})). The
hierarchy of $I(m,n,Z)$ is interrelated by partial fractions, differentiations
and partial integrations so that, in the case $k=0$
considered in Ref. \cite{Velt84},
all integrals can be derived from the basic formula for $I(2,1,1)$ which
itself can be written as an analytic expression.

For the problem considered here, we have to extend the formalism by
introducing a four momentum vector $k$ in the expressions (\ref{hierarchy})
which depends on the external momenta plus up to 2 additional
Feynman parameters. This implies no major changes in the Feynman
parametrization and evaluation of the four momentum integrals but markedly
complicates the subsequent Feynman parameter integrations: in general, we
obtain a four dimensional integral where, due to the $\eta$ instability
mentioned above, the integrand contains logarithmic singularities within the
parameter domain and is complex
for arbitrary values of the kinematic variables. It is essential to work
out the analytic structure of the integrand by performing at least two
integrations analytically; the remaining two--dimensional integral can then
be estimated numerically with reasonable accuracy. We will point out the
nontrivial steps of the procedure in the appendix, sect. C.

Because of the amount of algebra involved, it is essential to do the whole
calculation as systematically as possible. As far as the derivation of
integral expressions and the reduction and simplification of the
numerators $Z(l_1,l_2, q_1,q_2,P,x_1, \ldots, x_m)$ is concerned, we are
helped by the fact that the $\eta 3 \pi$--vertices always produce a constant
that factors out. The basic integrals $I(m,n,Z)$ can be checked independently
and various relations between them help to exclude errors in the derivation.
Finally, at least in the subthreshold region, all analytic one-- or
two--dimensional integral formulas can be easily checked on the computer.

Unless the loop corrections discussed so far, the overlapping diagrams
contribute to the amplitude part $B(s,t)$ as well as to  $A(s)$. This implies
that there is a contribution from these graphs even at $s=0$ where $A(s)$
has no influence (see Eq. (\ref{decayrate}) and Fig. \ref{endloopres}).
The analogy with the
pion production process where nonanalytical contributions were found
to be small in the energy range $2 m_{\pi} \leq \sqrt{s} \leq 400$ MeV
\cite{Gas94} can therefore not be used to argue that the effect of the
overlapping diagrams on the $\eta$--decay rate is negligible \cite{Belk95}.
Indeed, we find a
decay width from the overlapping pion diagrams alone that is comparable
to the one found for the factorizable diagrams
$\Gamma = 4.1 \times 10^{-4}$ eV, this changes the decay width
(\ref{widthfacloop}) to
\begin{equation}
 \Gamma^{L^6+O(p^4)+L^4+{\rm fact. + overl.}} (\eta \to \pi^0 \gamma \gamma) =
0.696
 \, {\rm eV}.
\label{widthfullloop} \end{equation}
Again, we might estimate an uncertainty of $\pm 0.04$ eV due to the kaon loops.
If -- as in the case of the ${\cal{O}}(p)^4$ one--loops -- the interference
of kaon loops and pion loops is equal, the prediction of chiral perturbation
theory to ${\cal{O}}(p)^6$ should be am $\eta$--decay width of about
0.76 eV.

As pointed out in Ref. \cite{Bij92}, the anomalous part of the ${\cal{L}}^4$
Lagrangian generates one--loop terms that are ${\cal{O}}(p)^8$
in the momentum expansion but of a size comparable to the ${\cal{O}}(p)^4$
loops because there is no G--parity or kaon mass suppression.
Adding these contributions (Eq. (27)/(28) of Ref. \cite{Bij92}) to
the amplitude of Eq. (\ref{widthfullloop}), we get the final result
\begin{equation}
 \Gamma^{L^6+O(p^4)+L^4+{\rm fact. + overl.}} (\eta \to \pi^0 \gamma \gamma)
 = 0.765
 \, {\rm eV}.
\label{widthend} \end{equation}

The coupling constants of the vector mesons used to determine the
${\cal{L}}^6$--constants are subject to errors.
Based on the uncertainties quoted in  \cite{Ng92} for the
$\omega$  and $\rho$ couplings, we calculate a relative error
of the tree level amplitude (\ref{treeamp}) of about 20 $\%$.
Together with the missing kaon loops, this implies an uncertainty to the
${\cal{O}}(p)^6$ amplitude of
\begin{equation}
 \Gamma^{L^6+O(p^4)+L^4+{\rm fact.}} (\eta \to \pi^0 \gamma \gamma) = 0.77
\pm 16 \, {\rm eV}.
\end{equation}
This is in close
agreement with the experimental value and the result of the quark box model
calculation of \cite{Ng93} but is markedly higher than the prediction of the
meson exchange model used to determine the Lagrangian.

Addition of the loop contributions to
the "all order fit" amplitude of Table II does not alter the interference
scheme described so far but would result in a total decay
width of
\begin{equation}
 \Gamma^{{\rm VMD + loops}} (\eta \to \pi^0 \gamma \gamma) = 0.439
\pm \, 9  {\rm eV}.
\label{VMDend} \end{equation}
The difference between Eqs. (\ref{widthend}) and (\ref{VMDend}) gives an
estimate of the uncertainty of the model induced by higher order terms in the
momentum  expansion. The various contributions to the decay width in both
the ${\cal{O}}(p)^6$ and the all order fit scheme are collected in
Table 3; the influence of the loops to the decay spectrum is shown in Fig.
\ref{endloopres}. The plot shows again that contributions to $B(s,t)$
affect the spectrum especially at low energy whereas the $A(s)$--contributions
are cut off at $s=0$.

\section{Conclusions}

We have considered the decay $\eta \to \pi^0 \gamma \gamma$
as an example for the application of the general chiral Lagrangian
${\cal{L}}^6$. A total of 6 structures contribute to the process; the
corresponding low--energy constants can be determined in a meson
exchange model if the analogous process $\gamma \gamma \to \pi^0 \pi^0$
is considered simultaneously. To be consistent at ${\cal{O}}(p)^6$, the
${\cal{L}}^6$ tree--level amplitude must be complemented by one-- and
two--loop amplitudes. The result is in agreement with the experimental
data for both the neutral pion production and the $\eta$--decay process.

We have given arguments for how to choose the relative signs of the
coupling constants in the meson exchange amplitude. As this amplitude
contains terms of ${\cal{O}}(p)^8$ and higher, there remains a sizeable
difference between the predictions of the full meson exchange model
and  ${\cal{O}}(p)^6$ chiral perturbation theory. The total decay
widths calculated in both models differ by a factor of 1.7
($\Gamma(VMD)=0.45$ eV, $\Gamma(\chi P T)=0.77$ eV. Furthermore,
unlike the meson exchange spectrum, the differential
width in chiral perturbation theory is small at low invariant energy $s$
and reaches a maximum at about 80 $\%$ of the maximum $s$ kinematically
allowed. An experiment measuring the
decay spectrum would discriminate between both predictions and thus
shed light on the details of the mechanism.

The loop contributions for the $\eta$--decay are all individually small.
However, as we know from the contribution of the ${\cal{O}}(p)^4$--loops
\cite{Bij92}, there can be sizeable interference effects with the
leading tree--level amplitude. After a detailed calculation we could show that
such interference effects occur (at the 5-10$\%$ level) in parts of the
amplitude but that the complete two--loop result is
suppressed by destructive interference,
in particular between the ${\cal{L}}^4$ kaon--loops and the
rest. The two kaon--loop amplitudes have not been calculated but rather
estimated to be sufficiently small. Our results confirm
the convergence of the chiral loop expansion.

The relative size of the two--loop contributions, in particular the
overlapping graphs displayed in Fig. \ref{overlapdiag} are not quite large
enough to justify the amount of work required for their calculation.
Nevertheless, we are confident that the strategy for solving the graphs
outlined in the appendix may be useful for related calculations.
In particular,
it would be worthwhile to treat the production amplitude $\gamma \gamma
\to \pi^0 \pi^0$ in SU(3)--chiral perturbation theory and compare our
formalism with the alternative dispersion--theoretical approach.

\section{Appendix: A Recipe for Solving Two--Loop Integrals in $\chi$PT }

\subsection{Restrictions, Remaining Vertices and Diagrams}

In this section, we develop a
method for solving the two--loop contributions for a two
neutral meson/two photon process applicable for equal masses of the
loop--mesons  and real photons but arbitrary external mesons.
(The general mass case would only introduce some modifications in the
analytic integrations of the overlapping diagrams treated in subsection C.)
The general diagrams contributing at ${\cal{O}}(p^6)$ in the chiral
expansion are shown in Fig. 2 of Ref. \cite{Gas94}. For the $\eta$--decay,
the diagrams involving ${\cal{L}}^4$ tree--level interactions are cancelled;
the only diagrams involving ${\cal{L}}^4$--terms are thus one--loop
diagrams with a single 4--meson vertex.

As ${\cal{L}}^2$ can only generate vertices with an even number of mesons,
the two--loop configurations we are left with are those displayed in
Fig. \ref{Configurations} where the squares denote ${\cal{L}}^2$--interactions
and photons are to be attached in all possible ways.
If we restrict ourselves to internal pions, the diagrams
contain each a vertex function $\Gamma_{\pi \pi \pi \eta}=$const. (see Eq.
(\ref{vertex}) below). By a
symmetric loop integration argument, one can show that this
excludes  the diagrams of the type (c) of Fig. \ref{Configurations}. We are
thus left with the factorizable diagrams (a) and (b) plus the overlapping
diagrams of type (d).

The vertex functions required can be calculated in a standard way by inserting
the expansion $U=1+{i \over F_{\pi}} \Phi - \ldots$ in the Lagrangian
${\cal{L}}^2$ and collecting all terms with a given number of meson and
photon fields. The procedure implies the trace of a product of up to six
$3 \times 3$--matrices and is therefore most conveniently done on the
computer. Neglecting the kaon sector, one obtains the following vertex
functions:
\begin{eqnarray} \Gamma^{\mu}_{\pi^+ \pi^-} & = & - i e (p_+-p_-); \qquad
\quad \quad \Gamma^{\mu \nu}_{\pi^+ \pi^-}  =  - 2 i e g^{\mu \nu}; \cr
\Gamma_{\pi^+ \pi^- \eta \pi^0} & =
& - {i \sqrt{2} \over 3 \sqrt{3} F_{\pi}^2} (\Delta
m_K)^2  \qquad
 \Gamma_{\pi^0 \pi^0 \eta \pi^0}
=  - {i \sqrt{2} \over  \sqrt{3} F_{\pi}^2} (\Delta
m_K)^2  \cr
\Gamma_{\pi^+ \pi^- \pi^0 \pi^0} & = &  {i \over 3 F_{\pi}^2 }\left(
2(p_1 p_2 + p_+ p_-)+(p_1+p_2)^2+m_{\pi}^2 \right) \cr
\Gamma_{\pi^+ \pi^- \pi^+ \pi^-} & = &  {i \over 3 F_{\pi}^2} \left(
p_1^2 + p_2^2 + p_+^2 + p_-^2-3(p_+-p_-)^2+2 m_{\pi}^2 \right) \cr
\Gamma^{\mu}_{\pi^+ \pi^- \pi^0 \pi^0} & = & - {2 i e\over 3 F_{\pi}^2 }
( p_+ - p_-)^{\mu}; \qquad \quad
\Gamma^{\mu}_{\pi^+ \pi^- \pi^+ \pi^-}  =   -{8 i \over 3 F_{\pi}^2}
( p_+ - p_-)^{\mu}; \cr
\Gamma^{\mu \nu}_{\pi^+ \pi^- \pi^0 \pi^0} & = & - {4 i e^2 \over 3 F_{\pi}^2}
 g^{\mu \nu}; \qquad \qquad \qquad \;
\Gamma^{\mu \nu}_{\pi^+ \pi^- \pi^+ \pi^-}  =  - {16 i e^2 \over 3 F_{\pi}^2}
 g^{\mu \nu}; \cr
\Gamma_{\pi^+ \pi^- \pi^0 \pi^0 \eta \pi^0} & = &
 {i \sqrt{2} \over 15 \sqrt{3} F_{\pi}^4} (\Delta m_K)^2;   \qquad
\Gamma_{\pi^+ \pi^- \pi^+ \pi^- \eta \pi^0}  =
 {i \sqrt{2} \over 10 \sqrt{3} F_{\pi}^4} (\Delta m_K)^2; \label{vertex}
\end{eqnarray}
All other vertex functions vanish. The conventions used for these expressions,
in particular the orientation of the four momenta,
are displayed in Fig. \ref{vertexpic}. Statistical factors for the case
that there are two or more identical particles have been included in Eqs.
(\ref{vertex}). For vertices involving two $\pi^+ \pi^-$ pairs, the momenta
$p_+$, $p_-$ have to be assigned symmetrically.
Note that there are no photon vertices
with neutral mesons only and that all the vertices involving
a single $\eta$--meson are generated by the mass term of the Lagrangian
${\cal{L}}^2$ and are constant. In analogy to the one--loop amplitude
\cite{Bij92}, the total internal pion two--loop amplitude
is G-parity suppressed.

\subsection{Factorizable Diagrams}

The Diagrams (a) and (b) of Fig. \ref{Configurations} contain no propagators
containing both loop momenta and therefore effectively factorize into a
product of two one--loop expressions. According to the scheme of dimensional
regularization \cite{Thooft72}, one--loop integrals contain singularities
linear in ${ 1 \over \epsilon} = {1 \over  2-{n \over 2}} $
where n is the continuous dimension
parameter. In order to retrieve all nonvanishing terms as $n \to 4$, we are
forced to develop the one--loop expressions to ${\cal{O}}(\epsilon)$.
The divergence of the resulting two--loop amplitude is at most quadratic
in ${1 \over \epsilon}$.

For clarity, we display the explicit expansions of the
simplest elementary one--loop diagrams. A tadpole loop
(see Fig. \ref{simplediag} (a)) generates \cite{Velt84}:
\begin{eqnarray} I_{tad}& \equiv &\int {d^4 l \over (2 \pi)^4} {1 \over [l]} =
-{i m^2 \over (4 \pi)^2} {\Gamma(2-{n \over 2}) \over 1-{n \over 2}}
\left({4 \pi \mu^2 \over m^2}\right)^{2-{n \over 2}}
 = {i m^2 \over (4 \pi)^2} \left(
{1 \over \epsilon} + \left[ 1-\gamma - \ln({m^2 \over 4 \pi \mu^2}) \right]
\right. \cr \mbox{} && \left.
+ \epsilon \left[ 1 + \delta - \gamma(1-\ln({m^2 \over 4 \pi \mu^2}) )
-\ln({m^2 \over 4 \pi \mu^2}) + {1 \over 2} \ln^2({m^2 \over 4 \pi \mu^2})
\right] + {\cal{O}}(\epsilon^2) \right), \label{tadpole} \end{eqnarray}
where we have denoted the propagator as $[l]=l^2-m^2+i \varepsilon$, expanded
the $\Gamma$--function
$$ \Gamma(\epsilon)={1 \over \epsilon} - \gamma + \epsilon \delta +
{\cal{O}}(\epsilon^2); \quad \delta={\pi^2 \over 12} + {\gamma^2 \over 2},$$
with the Euler constant $\gamma=0.577215665$ and introduced the renormalization
constant $\mu$. The small imaginary factor in the propagator will not be
written explicitly in the following but should be kept in mind in order to
retrieve the imaginary parts of the loop integrals. Also, in Eq.
(\ref{tadpole})
as well as in the following, the limits $n \to 4$ and equivalently
$\epsilon \to 0$ are always understood.

A symmetric bubble--diagram (Figure \ref{simplediag} (b))
with photon momenta $q_1+q_2$ where
$s=(q_1+q_2)^2$ yields after Feynman parametrization
\begin{eqnarray} I^s_{bub}& \equiv &  \int {d^4 l \over (2 \pi)^4} {1 \over
[l+{q_1+q_2 \over 2}] [l - {q_1+q_2 \over 2}] }  =  {i \over (4 \pi)^2}
 \int_0^1 dx  \Gamma(2-{n \over 2})
\left({4 \pi \mu^2 \over sx(1-x)-m^2}\right)^{2-{n \over 2}}
\cr & =& {i \over (4 \pi)^2} \left(
{1 \over \epsilon} + \left[ -\gamma - \ln({m^2 \over 4 \pi \mu^2}-G(s)) \right]
\right. \cr \mbox{} && \left.
+ \epsilon \left[  \delta + \gamma \ln({m^2 \over 4 \pi \mu^2})
+ {1 \over 2} \ln^2({m^2 \over 4 \pi \mu^2}+ \gamma G(s)+
\ln^2({m^2 \over 4 \pi \mu^2} G(s) + {1 \over 2}
G^2(s)
\right] + {\cal{O}}(\epsilon^2) \right). \label{bubble}  \end{eqnarray}
where the loop integrals
\begin{eqnarray} G(s)&=&\int_0^1 dx \ln(1-{s \over m^2}x(1-x)); \; \;
 \quad G(0)=0;
 \cr G^2(s)&=&\int_0^1 dx \ln^2(1-{s \over m^2}x(1-x)); \quad G^2(0)=0
\label{gloop} \end{eqnarray}
are analytic functions cut along the real axis as $s > 4 m^2$. The
integrals G(s) and
\begin{equation} F(s)=\int_0^1 {dx  \over x} \ln(1-{s \over m^2}x(1-x));
\quad F(0)=0
\end{equation}
can be easily reduced to analytic expressions \cite{Hol88,Gas94}; moreover,
loop integrals with a single logarithm but additional factors $x, \, x^2$
etc. can be expressed in terms of $G(s)$ and $F(s)$.
The loop functions of the form $G^2(s), \, F^2(s)$ etc.
can be expressed in terms of polylogarithms \cite{Lev81} or
be solved numerically.

The scalar vertex diagram (scalar meaning numerator equal to 1, see Fig.
\ref{simplediag} (c)) is finite and written as
\begin{eqnarray} I_{\Delta}& = &  \int {d^4 l \over (2 \pi)^4} {1 \over
[l] [l-q_1] [l+q_2] }  =  {i \over (4 \pi)^2}
 \int_0^1 dx_1 \int_0^{1-x_1} dx_2  \Gamma(3-{n \over 2})
{(4 \pi \mu^2)^{2-{n \over 2}}
  \over (sx_1 x_2 -m^2)^{3-{n \over 2}} }
\cr & =& {i \over (4 \pi)^2}
 \int_0^1 dx_1 \int_0^{1-x_1} dx_2 { 1 + \epsilon \left(
-\gamma - \ln({m^2 \over 4 \pi \mu^2})- \ln(1-{s \over m^2} x_1 x_2) \right)
\over s x_1 x_2 - m^2   } \label{triangle} \end{eqnarray}

The scalar integrals considered so far can be easily generalized to the
case where loop momenta appear in the numerator of the integrand. For
the case of a linear momentum in the numerator, the formula for dimensional
regularization are applicable. Integrals with quadratic and higher order
loop momenta in the numerator are reduced to simpler integrals
with formula such as
\begin{equation} \int {d^4 l \over (2 \pi)^4}
{4 l^{\mu} l^{\nu}- {4 \over n} g^{\mu \nu} l^2 \over
(l^2 + 2 k l - m^2 )^\alpha }  =  {i \over 16 \pi^{n \over 2} }
{4 k^{\mu} k^{\nu}- {4 \over n} g^{\mu \nu} k^2 \over
(-k^2 - m^2 )^{\alpha-{n \over 2}} }  \end{equation}

The vertex and box diagrams appearing in the amplitude generate further loop
functions of the types
$$   \int_0^1 dx_1 \int_0^{1-x_1} dx_2 {x_1^m x_2^n \over s x_1 x_2 -m^2};
\qquad \int_0^1 dx_1 \int_0^{1-x_1} dx_2 {x_1^m x_2^n
\ln(1-{s \over m^2}x_1 x_2)   \over s x_1 x_2 -m^2};
$$ $$
\int_0^1 dx_1 \int_0^{1-x_1} dx_2 {x_1^m x_2^n  \over (s x_1 x_2 -m^2)^2};
\qquad \int_0^1 dx_1 \int_0^{1-x_1} dx_2 {x_1^m x_2^n
\ln(1-{s \over m^2}x_1 x_2)   \over (s x_1 x_2 -m^2)^2};
$$
with powers $0 \leq m,n \leq 3$. The loop functions are symmetric in m and n.
In all cases, the $x_2$--integration is
easy to perform analytically; partial integration and simple algebraic
manipulations allow to express everything  in terms of
$F(s), \; G(s), \, F^2(s)$ (defined  as $F(s)$ but with a quadratic logarithm
in the integrand) and $G^2(s)$.
Due to the similarity of the loop functions, we find that the factorizable
amplitude has a unique s--channel threshold $s=4 m^2$.

Let us illustrate the method in a simple example, the 6-meson vertex
diagram of Fig. \ref{Configurations} (a). In order to get a nonvanishing
amplitude, both photon lines have to be attached to one charged meson loop;
the second loop can be charged or neutral. From the vertex expressions
in Eq. (\ref{vertex}), one finds that the amplitude can be written as
\begin{equation} M^{\mu \nu} =
 - {3 \over 2}{i \sqrt{2} e^2 (\Delta m_K)^2 \over 9
\sqrt{3} F_{\pi}^4 } \int {d^4 l_1 \over (2 \pi)^4} {d^4 l_2 \over (2 \pi)^4}
{(2 l_1-q_1)^{\mu}(2 l_1 + q_2)^{\nu} - g^{\mu \nu}(l_1^2 -m^2) \over
[l_1][l_1-q_1][l_1+q_2][l_2] } + (q_1,\mu \leftrightarrow q_2,\nu)
\label{sixpi} \end{equation}
This is gauge invariant because of
\begin{eqnarray} q_{1 \mu} M^{\mu \nu} & \sim & \int {d^4 l_1 \over (2 \pi)^4}
{q_{1 \mu} \left[ (2 l_1-q_1)^{\mu}(2 l_1 + q_2)^{\nu} -
 g^{\mu \nu}(l_1^2 -m^2)  \right] \over
[l_1][l_1-q_1][l_1+q_2] } \cr
& = &  \int {d^4 l_1 \over (2 \pi)^4}
{2 l_1^{\nu} \over [l_1-{q_1+q_2 \over 2}][l_1+{q_1+q_2 \over 2}] }
 -   \int {d^4 l_1 \over (2 \pi)^4}
{2 l_1^{\nu} \over [l_1-{q_2 \over 2}][l_1+{q_2 \over 2}] } = 0.
 \end{eqnarray}
With the expressions (\ref{tadpole}), (\ref{bubble}) and (\ref{triangle})
for the elementary diagrams, the amplitude (\ref{sixpi}) reads
\begin{eqnarray} M^{\mu \nu} &=& - {i \sqrt{2} e^2 (\Delta m_K)^2 \over 3
\sqrt{3} F_{\pi}^4 } \left( g^{\mu \nu} {s \over 2} - q_2^{\mu} q_1^{\nu}
\right)  I_{tad} \times \cr \mbox{} && \times  {i \over (4 \pi)^2}
 \int_0^1 dx_1 \int_0^{1-x_1} dx_2 {x_1 x_2 \left[ 1 + \epsilon \left(
-\gamma - \ln({m^2 \over 4 \pi \mu^2})- \ln(1-{s \over m^2} x_1 x_2) \right)
\right] \over s x_1 x_2 - m^2   }. \end{eqnarray}
The loop integrals are readily solved and yield
\begin{eqnarray}
 \int_0^1 dx_1 \int_0^{1-x_1} dx_2 {x_1 x_2 \over s x_1 x_2 - m^2 }
& = &  {1 \over s} \left( {1 \over 2} + {m^2 \over s} F(s) \right) \cr
 \int_0^1 dx_1 \int_0^{1-x_1} dx_2 {x_1 x_2 \ln(1-{s \over m^2} x_1 x_2)
\over s x_1 x_2 - m^2 }
& = &  {1 \over s} \left( -{1 \over 2} - {m^2 \over s} F(s)
+{1 \over 2} G(s) + {m^2 \over 2 s} F^2(s) \right) \cr
\end{eqnarray}
The 6-meson diagrams contribute to the form factor A(s,t) of Eq. (\ref{eich})
only. The same is true for all factorizable diagrams.

\subsection{Overlapping Diagrams}

The overlapping diagrams derived from Fig. \ref{Configurations} (d)
translate into sums of integrals of the type
\begin{eqnarray}  I(m,n,Z) & = &
\int_0^1 dx_1 \cdots \int_0^{1-x1-\ldots-x_{n-1}}dx_n
\int_0^1 dx_{n+1} \cdots \int_0^{1-x_{n+1}-\ldots-x_{m-1}}dx_m
\cr && \mbox{} \quad \times
\int {d^4 l_1 \over (2 \pi)^4} {d^4 l_2 \over (2 \pi)^4}
{Z(l_1,l_2,q_1,q_2,P,x_1,\ldots ,x_m) \over
[l_1]^n[l_2]^m[l_2-l_1-k(x_1,\ldots ,x_m)]},
\label{hierarchy} \end{eqnarray}
where $\quad  1 \leq n, \, m-n \leq 3; \quad m \leq 4$.
Here, we have used Feynman parameters in order to put together propagators
with equal loop momenta and obtain more symmetric expressions.
The numerators $Z$ can be largely simplified by factoring out $l$--dependent
terms; the remainder is at most an expression cubic in the loop momenta
$l^{\mu} l^{\nu} l^{\rho}$.
With further Feynman parametrization,
all the integrals $I(m,n,Z)$ can be cast in a  form that makes the analytic
structure of the loop integrals explicit and leaves a two--dimensional
numerical integration over a well--behaved, complex function.
We demonstrate the method
for the scalar integrals $I(2,1,1)$, $I(2,2,1)$ and $I(3,1,1)$.

The integral $I(2,1,1)$ corresponds to the scalar part of the diagram
Fig. \ref{overlapdiag} (c). The analytic case where $k=0$ has been
considered in \cite{Velt84}. In extension to the method outlined there,
we apply subsequent feynman parametrization and integration over
the loop momenta and write
\begin{eqnarray} I(2,1,1) & \equiv & \int_0^1 dx
\int {d^4 l_1 \over (2 \pi)^4} {d^4 l_2 \over (2 \pi)^4}
{1 \over
[l_1]^2[l_2][l_2-l_1-k(x)]} \cr
& = &  \int_0^1 dx
\int {d^4 l_1 \over (2 \pi)^4} {i \over (4 \pi)^2 }
\int_0^1 dz {\Gamma(2-{n \over 2})(4 \pi \mu^2)^{2-{n \over 2}}
 \over \left( z(1-z) \right)^{2-{n \over 2}}
}  {1 \over
[l_1]^2 \left( (l_1+k(x))^2 - {m^2 \over z(1-z)} \right)^{2-{n \over 2}} } \cr
& = &  \int_0^1 dx
{-1 \over (4 \pi)^4}
\int_0^1 dz {1 \over \epsilon } \left( {(4 \pi \mu^2)^2 \over
z(1-z)} \right)^{2-{n \over 2}}
\int_0^1 dy \, y (1-y)^{2-{n\over 2}} \cr & & \mbox{} \times \quad \left\{
{ \left( k^2-{m^2 \over z(1-z)}-2k^2(1-y)+k^2 (1-y)^2 \right) \Gamma(5-n) \over
\left(  k^2 y (1-y) - m^2 y - {m^2 (1-y) \over z(1-z)} \right)^{5-n} }
\right. \cr & & \mbox{} \left.
\qquad \qquad + {n \over 2} {\Gamma(4-n) \over
\left( k^2 y (1-y) - m^2 y - {m^2 (1-y) \over z(1-z)} \right)^{4-n} }
\right\} \end{eqnarray}
After expanding $(1-y)^{2-{n \over 2}} {n \over 2}$ in powers of
$\epsilon=2-{n \over 2}$, the second term in this expression can be partially
integrated over y. One obtains a sum of terms where the z-integration
plus some trivial x- and y-integrations can be done analytically. The final
result can be cast in the form
\begin{eqnarray} I(2,1,1) & = & {-1 \over (4 \pi)^4} \left\{ {1 \over 2
\epsilon^2} + { 1 \over 2 \epsilon} \left( 1-2 \gamma -
 2 \ln \left( {m^2 \over 4 \pi \mu^2} \right) \right) \right.
\cr  && \mbox{} \left. \quad + {\pi^2 \over 12} + {5 \over 2}
- \gamma + \gamma^2 -(1-2 \gamma) \ln \left({m^2 \over 4 \pi \mu^2} \right)
  +  2 \ln^2 \left({m^2 \over 4 \pi \mu^2} \right) \right.
\cr  && \mbox{} \left. \quad
+ \int_0^1 dy G(a) - \int_0^1 {dy \over y } \left( G(a) + \ln(y) +2 \right)
\right\}; \cr
a & \equiv & {k^2(x) y (1-y) -m^2 (1-y) \over m^2 y }
\label{simplescal} \end{eqnarray}
with the loop function $G$ from Eq. (\ref{gloop}). We note that
for the $\eta$--decay diagram Fig. \ref{overlapdiag} (c), the external
momentum variable can be written as $k=P-q_1 x$ . Owing to the properties of
$G$ and the ratio of the $\eta$-- and $\pi$--masses, the integrand in Eq.
\ref{simplescal} is a complex function for arbitrary $s \geq 0$.
The integral is finite with a logarithmic end point singularity in the
real part of the integrand as $y \to 0$.

Feynman integrals with a numerator depending on the loop momenta can be
solved analogously. One obtains
\begin{eqnarray}
I(2,1,l_1^{\mu}) & \equiv &
 \int_0^1 dx
\int {d^4 l_1 \over (2 \pi)^4} {d^4 l_2 \over (2 \pi)^4}
{l_1^{\mu} \over
[l_1]^2[l_2][l_2-l_1-k(x)]} \cr
& = & {-1 \over (4 \pi)^4} \int_0^1 dx (-k(x))^{\mu} \left\{ {1 \over 4
\epsilon} - {1 \over 8}
- {\gamma \over 2} - {1 \over 2} \ln \left({m^2 \over 4 \pi \mu^2} \right)
+ \int_0^1 dy (1-y) G(a)  \right\}; \cr
I(2,1,\Delta l_1^{\mu \nu}) & \equiv &
 \int_0^1 dx
\int {d^4 l_1 \over (2 \pi)^4} {d^4 l_2 \over (2 \pi)^4}
{4 l_1^{\mu} l_1^{\nu} - {4 \over n} l_1^2 g^{\mu \nu} \over
[l_1]^2[l_2][l_2-l_1-k(x)]} \cr
& = & {-1 \over (4 \pi)^4} \int_0^1 dx (4 k(x)^{\mu} k(x)^{\nu} - {4 \over n}
k(x)^2 g^{\mu \nu})
\cr  && \mbox{} \quad \times
\left\{ {1 \over 12
\epsilon} - ({1 \over 24}+ {1 \over 18})
- {\gamma \over 6} - {1 \over 6} \ln \left({m^2 \over 4 \pi \mu^2} \right)
+ \int_0^1 dy \, y (1-y) G(a)  \right\}; \cr
\end{eqnarray}
The last equation shows how to reduce integrals with higher powers of loop
momenta in the numerator to the integral class $I(1,1,Z)$. The latter
integral can in turn be related to $I(2,1,Z)$ by means of the
'partial' operation \cite{Thooft72,Hoog85}.

The 'master' diagram Fig. \ref{overlapdiag} (b) leads to a scalar integral
$I(2,2,1)$ with $k(x_1,x_2)=P-q_1 x_1-q_2 x_2$. The integral is {\it finite} so
that we don't have to worry about the regularization procedure:
\begin{eqnarray} I(2,2,1) & \equiv & \int_0^1 dx_1 \int_0^1 dx_2
\int {d^4 l_1 \over (2 \pi)^4} {d^4 l_2 \over (2 \pi)^4}
{1 \over
[l_1]^2[l_2]^2 [l_2-l_1-k(x_1,x_2)]} \cr
& = & {-1 \over (4 \pi)^4}
 \int_0^1 dx_1 \int_0^1 dx_2
\int_0^1 dz \int_0^1 dy \, {z y \over \left( k^2 y (1-y)- m^2 y \right) z (1-z)
-m^2 (1-y) } \cr
& = & {-1 \over (4 \pi)^4}
 \int_0^1 dy_1 \int_0^{1-y_1} dy_2
\int_0^1 dz_1 \int_0^{1-z_1} dz_2 \, {1 \over \tilde r^2 z_1 z_2 + \tilde s^2
z_1 (1-z_1) - \tilde m^2  } \cr
& = & {-1 \over (4 \pi)^4}
 \int_0^1 dy_1 \int_0^{1-y_1} dy_2 \;
{1 \over \tilde r^2} \left( F\left({\tilde r^2 + \tilde s^2 \over \tilde m^2}
\right) -F\left({\tilde s^2 \over \tilde m^2}\right) \right),
\label{master} \end{eqnarray}
where $\tilde r^2, \, \tilde s^2$ and $\tilde m^2$ depend on the Feynman
parameters $y_1, \, y_2$ and the invariant kinematic variables
\begin{eqnarray}
\tilde r^2 & \equiv & s y_1 y_2 - (m^2 - u) y_1 (1-y_1),
\cr
\tilde s^2 & \equiv & - (m^2 - t) y_1 y_2 - m^2 (1-y_1)^2 ,
\cr
\tilde m^2 & \equiv &  m^2 y_1 .
\nonumber \end{eqnarray}

The analytic integration is largely simplified by the fact that both
photons are real ($q_1^2 = q_2^2 = 0$) and the masses in the propagators
are identical. The general mass case can still be treated in the same way,
but with polylogarithms occurring from the analytical $z$--integration.

The integral $I(2,2,1)$ is linked to $I(2,1,1)$ by
\begin{equation} I(2,2,1)=\lim_{M^2 \to m_2} {\partial \over \partial M^2}
 \int_0^1 dx_1 \int_0^1 dx_2
\int {d^4 l_1 \over (2 \pi)^4} {d^4 l_2 \over (2 \pi)^4}
{1 \over
[l_1]^2 (l_2^2-M^2) [l_2-l_1-k(x_1,x_2)]}.
\label{deriv} \end{equation}
Relations such as Eq. (\ref{deriv}) can be used to check the calculation.

For an accurate numerical estimate of two--loop Feynman integrals,
it is crucial to evaluate the first complex integration analytically.
Due to the $x$--parametrization of loop momenta and the presence of
$l$--dependent terms
in the numerator $Z$, the master diagram contains integrals of the form
\begin{equation} I^{M}(l,m,n)=\int_0^1 dz_1 \int_0^{1-z_1} dz_2
{z_1^l z_2^m \over (1-z_1)^n} {1 \over a z_1 z_2 + b z_1 (1-z_1 )-c};
\quad 0 \leq l,m,n \leq 3. \label{parameterint} \end{equation}
The case $m>0$ leads to cancelling singular terms after the first integration
and requires formula such as
\begin{eqnarray}
& \int_0^1 {dz \over z} \left\{ 1 + {c \over a z}\left( \ln \left(1-
{a+b \over c} z(1-z) \right) - \ln \left(1- {b \over c} z (1-z) \right) \right)
 \right\}& \cr &\mbox{} \qquad \qquad
= 1 + G\left({a+b \over c}\right)+ {b \over a}
\left( G\left({a+b \over c}\right)
-G\left({b \over c}\right) \right) &
\nonumber
\end{eqnarray}
that are conveniently derived by formal partial integration of the
logarithmic terms and subsequent rearrangement of the integrands
keeping in mind the infinitesimal complex increment so that, e.\ g.
$$
\int_0^1 {dz \over a z^2 -az + 1 }  =  {2 \over 4-a } \left( G(a)+2 \right).
\nonumber $$
It turns out that all integrals $I^M(l,m,n)$ can be expressed in terms
of the loop functions $F$ and $G$. As in Eq. (\ref{simplescal}),
we are left with a two--dimensional integral over a complex function that
can be easily done numerically.

The diagram Fig. \ref{overlapdiag} (a) yields $k=P-q_1(1-x_1)-q_2x_2$
and the modified integral
\begin{eqnarray} \tilde I(3,1,1) & \equiv & \int_0^1 dx_1 \int_0^{1-x_1} dx_2
\int {d^4 l_1 \over (2 \pi)^4} {d^4 l_2 \over (2 \pi)^4}
{1 \over
(l_1^2+s x_1 x_2 -m^2)^3[l_2] [l_2-l_1-k(x_1,x_2)]} \cr
& = & {-1 \over (4 \pi)^4}
 \int_0^1 dx_1 \int_0^{1-x_1} dx_2
\int_0^1 dz {1 \over \epsilon } \left( {(4 \pi \mu^2)^2 \over
z(1-z)} \right)^{2-{n \over 2}}
\int_0^1 dy \,
y^2 (1-y)^{2-{n\over 2}} \cr & & \mbox{} \times \quad \left\{
{ \left( k^2 y -{m^2 \over z(1-z)} \right) \Gamma(6-n) \over
\left(  k^2 y (1-y) +(s x_1 x_2- m^2) y - {m^2 (1-y) \over z(1-z)}
\right)^{6-n} }
\right. \cr & & \mbox{} \left.
\qquad \qquad + {n \over 2} {\Gamma(5-n) \over
\left( k^2 y (1-y) + (s x_1 x_2 - m^2) y -
{m^2 (1-y) \over z(1-z)} \right)^{5-n} }
\right\} \end{eqnarray}
At this point, it is convenient to write the denominators as
\begin{eqnarray} & \left[ k^2 y (1-y) + (s x_1 x_2 - m^2) y -
{m^2 (1-y) \over z(1-z)} \right]^{-t}&  \cr & \mbox{} \qquad \qquad
= {1 \over y (t-1)}
\lim_{M^2-m^2} {\partial \over \partial
M^2}\left[ k^2 y (1-y) + (s x_1 x_2 - M^2) y -
{m^2 (1-y) \over z(1-z)} \right]^{-t+1},&  \nonumber \end{eqnarray}
partially integrate the second term as in Eq. (\ref{simplescal}), expand
and rearrange all terms and finally reperform the partial derivation.
The result can be represented as
\begin{eqnarray} \tilde I(3,1,1) & = & {-1 \over (4 \pi)^4}
\int_0^1 dx_1 \int_0^{1-x_1} dx_2  \left\{ \int_0^1 dz \int_0^1 {dy \over 1-y}
\right. \cr && \mbox{} \left. \left( {y^2 \over k^2 y(1-y) + (s x_1 x_2 -m^2)
y - {m^2(1-y) \over z (1-z)} }  - { 1 \over sx_1 x_2 -m^2} \right)
\right. \cr && \mbox{} \left.  \qquad \qquad + \int_0^1 dz
{1 \over \epsilon } \left( {(4 \pi \mu^2)^2 \over
z(1-z)} \right)^{2-{n \over 2}} {\Gamma(5-n) \over (s x_1 x_2 -m^2)^{5-n} }
\right\} . \label{standard} \end{eqnarray}
Note that $\tilde I(3,1,1)$ contains simple s--channel threshold terms
in $s x_1 x_2 - m^2$ as well as an anomalous threshold term. The cancellation
of the parameter
singularities in the first two contributions to Eq. (\ref{standard})
can be used as a test for the analytic integration.
The $x$-integrations in the s-threshold terms can be performed
according to subsection B. In the first term of Eq. (\ref{standard}), the
$x_2$-- and $z$--integrations can be done analytically yielding a form
analogous to Eq. (\ref{master}). As in the case of the master diagram
(see Eq. (\ref{parameterint})), the diagram \ref{overlapdiag} (a)
involves a whole set of analytic parameter integrals.

In analogy to the integrals of the class $I(2,1,Z)$, singularities may cancel
in the presence of $l$--dependent numerators. For instance, we obtain
\begin{eqnarray}
\tilde I(3,1,l_1^{\mu}) & \equiv &
 \int_0^1 dx_1 \int_0^{1-x_1} dx_2
\int {d^4 l_1 \over (2 \pi)^4} {d^4 l_2 \over (2 \pi)^4}
{l_1^{\mu} \over
(l_1^2 + s x_1 x_2 -m^2)^3[l_2][l_2-l_1-k(x)]} \cr
& = & {-1 \over (4 \pi)^4}
 \int_0^1 dx_1 \int_0^{1-x_1} dx_2
\int_0^1 dz \int_0^1 dy \,
{ y^2 (-k(x_1, x_2))^{\mu} \over
k^2 y (1-y) +(s x_1 x_2- m^2) y - {m^2 (1-y)  \over z(1-z)} }
\cr
\tilde I(3,1,\Delta l_1^{\mu \nu}) & \equiv &
 \int_0^1 dx_1 \int_0^{1-x_1} dx_2
\int {d^4 l_1 \over (2 \pi)^4} {d^4 l_2 \over (2 \pi)^4}
{4 l_1^{\mu}  l_1^{\nu} - {4 \over n} l_1^2 g^{\mu \nu}  \over
(l_1^2 + s x_1 x_2 -m^2)^3[l_2][l_2-l_1-k(x)]} \cr
& = & {-1 \over (4 \pi)^4}
 \int_0^1 dx_1 \int_0^{1-x_1} dx_2
\int_0^1 dz \int_0^1 dy \,
{ y^2 (1-y) (k^{\mu} k^{\nu} - g^{\mu \nu} k^2 ) \over
k^2 y (1-y) +(s x_1 x_2- m^2) y - {m^2 (1-y)  \over z(1-z)} }
\end{eqnarray}
This closes the list of basic integrals required for the type of processes
we are considering.

\bigskip \bigskip
I wish to thank Prof. H.\ W.\ Fearing for drawing my attention to this
topic and for valuable discussions.
I also thank S.\ Scherer, A.\ Schreiber and M.\ Welsh for discussions
concerning the subject.

This work was supported in part by a grant from the Natural Sciences and
Engineering Research Council of Canada.

\pagebreak
\begin{table}\centering
\begin{tabular}
      {@{~~~~~~}c@{~~~~~~}@{~~~~~~}c@{~~~~~~}c@{~~~~~~}
                             @{~~~~~~}c@{~~~~~~}c@{~~~~~~}}
  \\

{\mbox{\rule[-.40cm]{0cm}{0.9cm}}} Meson  &
\mbox{\rule[-.4cm]{0cm}{0.9cm} m [MeV]} &
\mbox{\rule[-.4cm]{0cm}{0.9cm}$ G_{\pi} \, [{\rm GeV}^{-2}]$}  &
\mbox{\rule[-.4cm]{0cm}{0.9cm}$ G_{\eta} \, [{\rm GeV}^{-2}]$}  &
\mbox{\rule[-.4cm]{0cm}{0.9cm}$ \Gamma_{\eta} [{\rm eV}]$}
  \\
 \hline     & & & &  \\
$ \rho $
 & 768 & 0.084 & 0.17  & 0.11
\\
$ \omega $
 & 782 & 0.49  & 0.098 & 0.28
\\
$ \Phi$
 & 1019 & 0.018  & 0.0089 &  0.29
\\
$ b_1 $
 & 1232 & 0.39 & 0.18  &  0.32
\\
$ a_0 $
 & 983  & 0 & -0.018/-0.022  &  0.35
\\
$ a_2 $
 & 1318 & 0 & 0.216 &  0.37
\\
$ f_0 $
 & 974 &  -0.007/-0.009 & 0 & -
\\
$ f_2 $
 & 1275 &  1.28 & 0  &  -
\\
\end{tabular}
\medskip
\caption[]{Meson parameters used for the meson exchange amplitudes of
Eqs. (\ref{etavmd}) -- (\ref{etascal}).
For the scalar mesons, both coupling constants $G^{Sd}$ and $G^{Sm}$
are given. The last column contains the (cumulative) total $\eta$--decay width.
\label{VMDcon} }

\end{table}
\begin{table}\centering
\begin{tabular}
      {@{~~~~~}c@{~~~~~} @{~~~~~}c@{~~~~~} @{~~~~~}c@{~~~~~}
    @{~~~~~}c@{~~~~~} @{~~~~~}c@{~~~~~} @{~~~~~}c@{~~~~~} @{~~~~~}c@{~~~~~} }
  \\

\mbox{\rule[-.40cm]{0cm}{0.9cm}} Model &
\mbox{\rule[-.4cm]{0cm}{0.9cm} $d_1$ } &
\mbox{\rule[-.4cm]{0cm}{0.9cm}$ d_4$}  &
\mbox{\rule[-.4cm]{0cm}{0.9cm}$ d_2$}  &
\mbox{\rule[-.4cm]{0cm}{0.9cm}$ d_5$}  &
\mbox{\rule[-.4cm]{0cm}{0.9cm}$ d_3$}  &
\mbox{\rule[-.4cm]{0cm}{0.9cm}$ d_6$}
  \\
 \hline     & & & & & & \\
$ {\cal{O}}(p^6) $
 & -1.60  &   0.654  &  3.80  & -1.93  &  38.9  & -16.1
\\
 All Order
 & 0.108  &    1.64  &  0.0486  &  -4.13  &  40.3  & -16.7
\\
\end{tabular}
\medskip
\caption[]{${\cal{L}}^6$ low--energy constants determined in two different
fit schemes (see text). All constants are in units of $10^{-3}$ GeV$^{-2}$.
The relation between $d_1, \ldots , d_6$ and the low energy constants
of Ref. \cite{Sche94} are given in Eq. (\ref{bconstants}) \label{l6const} }

\end{table}
\begin{table}\centering
\begin{tabular}
 {@{~~~}c@{~~~} @{~~~}c@{~~~} @{~~~}c@{~~~} @{~~~}c@{~~~}
    @{~~~}c@{~~~} @{~~~}c@{~~~} @{~~~}c@{~~~} @{~~~}c@{~~~} }
  \\

\mbox{\rule[-.40cm]{0cm}{0.9cm}} Model &
\mbox{\rule[-.4cm]{0cm}{0.9cm} ${\cal{L}}^6$ } &
\mbox{\rule[-.4cm]{0cm}{0.9cm}$ {\cal{O}}(p)^4$ }  &
\mbox{\rule[-.4cm]{0cm}{0.9cm}$ {\cal{L}}^4 $ }  &
\mbox{\rule[-.4cm]{0cm}{0.9cm} fact. }  &
\mbox{\rule[-.4cm]{0cm}{0.9cm} overl. }  &
\mbox{\rule[-.4cm]{0cm}{0.9cm} anom. }  &
\mbox{\rule[-.4cm]{0cm}{0.9cm} (K--loops) }
  \\
 \hline   & & & & & & & \\
$ {\cal{O}}(p^6) $
 & 0.652  &   0.733  &  0.673  & 0.682  &  0.696  & 0.765 & (0.83)
\\
 All Order
 & 0.371  &   0.411  &  0.381  &  0.383 &  0.386  & 0.439 & (0.46)
\\
\end{tabular}
\medskip
\caption[]{Cumulative contributions of the loop contributions to the total
$\eta$--decay width. The ${\cal{L}}^6$--fit parameters are those
given in Table II, the loop corrections are denoted as in the text. The last
column contains the estimate of the missing ${\cal{O}}(p)^6$ kaon loops.
\label{loopcont} }

\end{table}

\pagebreak

\begin{figure}
\begin{picture}(10,6)(0,0)
\end{picture}

\caption[]{\label{pivmd} The $\gamma \gamma \to \pi^0 \pi^0$
spectrum as measured by the
Crystal Ball Collaboration \cite{Mar90}. The curves are results of chiral
perturbation theory complemented by a ${\cal{O}}(p^6)$ meson exchange
amplitude: Pure chiral one-- and two--loop contributions in the SU(2)--model
(Eqs. (7.2) and (7.7) of Ref. \cite{Gas94}) (dashed), loop amplitude
plus full meson exchange model Eqs. (\ref{etavmd}) -- (\ref{etascal})
(dotted) and loop amplitude with meson exchange model but without axial
mesons Eq. (\ref{etaaxi}). }
\end{figure}

\begin{figure}
\begin{picture}(10,6)(0,0)
\end{picture}

\caption[]{\label{etafit} Predictions for the $\eta \to  \pi^0 \gamma \gamma$
spectrum. The predicion of the full meson exchange model (solid)
and its best fit according to the amplitude Eq. (\ref{etaampli}) (dotted)
yields
a total decay width of $\Gamma_{tot}=0.37$ eV. For comparison, the predictions
of the ${\cal{O}}(p^6)$ fit (dashed, $\Gamma_{tot}=0.65$ eV) and the
2-parameter fit according to \cite{Bij92}, Eq. (\ref{Bijlag}) (short dashed,
$\Gamma_{tot}=0.18$ eV) are also shown. Loop corrections are not taken into
account.}
\end{figure}

\begin{figure}
\begin{picture}(10,6)(0,0)
\end{picture}

\caption[]{\label{pifit} $\gamma \gamma \to \pi^0 \pi^0$
cross section from \cite{Mar90} and predictions from chiral perturbation
theory. \ref{etafit}. The results have been obtained with the
chiral one-- and two--loop contributions
(Eqs. (7.2) and (7.7) of Ref. \cite{Gas94}) complemented
with the ${\cal{O}}(p^6)$--contributions as in Fig. \ref{etafit}.
Axial vector mesons are not taken into account. }
\end{figure}

\begin{figure}
\begin{picture}(10,6)(0,0)
\end{picture}

\caption[]{\label{Configurations} The two types of factorizable two--loop
diagrams
contributing to the $\eta \to \pi^0 \gamma \gamma$ amplitude and considered
in this paper.   }
\end{figure}

\begin{figure}
\begin{picture}(10,6)(0,0)
\end{picture}
\caption[]{\label{vertexpic} Vertices and vertex functions
$\Gamma^{\mu}_{\pi^+ \pi^-}$, $\Gamma^{\mu \nu}_{\pi^0 \pi^0 \pi^+ \pi^-}$
and $\Gamma_{\pi^+ \pi^- \pi^0 \pi^0 \pi^0 \eta}$,
as used in Eq. (\ref{vertex}). }

\end{figure}

\begin{figure}
\begin{picture}(10,6)(0,0)
\end{picture}

\caption[]{\label{simplediag} Basic one--loop diagrams calculated in Appendix,
subsection B: Tadpole (a), bubble (b) and vertex (c) diagram.
}
\end{figure}

\begin{figure}
\begin{picture}(10,6)(0,0)
\end{picture}

\caption[]{\label{overlapdiag} The four types of overlapping diagrams occuring
in the process $\eta \to \pi^0 \gamma \gamma$.
}
\end{figure}

\begin{figure}
\begin{picture}(10,6)(0,0)
\end{picture}

\caption[]{\label{endloopres} Effect of the loop contributions on the
$\eta$--decay spectrum. The ${\cal{L}}^6$ tree--level results according to
Table III, first column are solid (compare Fig. \ref{etafit}),
the dashed lines include
the ${\cal{O}}(p)^4$ one--loop result (Table III, second column),
the dotted lines correspond to
the full result (Table III, second last column).
}
\end{figure}

\end{document}